\documentclass{aa}  

\usepackage{graphicx}
\usepackage{txfonts}
\begin{document}

      \title{A Forming Wide Polar Ring Galaxy at $z\sim0.05$ in the VST Deep Field of the Fornax Cluster\thanks{This work is based on observations taken at the ESO La Silla Paranal Observatory within the VST GTO Program ID 088.B-4012(A). }}

   \author{E. Iodice\inst{1}
          \and M. Capaccioli\inst{1,2}
          \and M. Spavone\inst{1}
          \and   N.R. Napolitano\inst{1}
          \and   A. Grado\inst{1}
	   \and    L. Limatola\inst{1}          
          \and M. Cantiello\inst{3}
          \and P. Schipani\inst{1}
}

   \institute{INAF-Osservatorio Astronomico di Capodimonte, via Moiariello 16, I-80131 Naples, Italy
              \email{iodice@na.astro.it}
                       \and
 	      Dipartimento di Fisica, Universit\'a di Napoli "Federico II'', via Cintia 21,  I-80126 Naples, Italy
              \and
 	      INAF - Osservatorio Astronomico di Teramo, via Maggini,  I-64100 Teramo, Italy
              }

   \date{Received ....; accepted ...}

 
  \abstract{
We present the first deep photometry of  a good candidate for a forming polar ring galaxy at  redshift $z\simeq0.05$.  This object, named FCSS J033710.0-354727, is a background galaxy in the VST deep field of the Fornax cluster. 
The deep exposures combined with the high angular resolution of the OmegaCAM at VST allow us to carry out the first detailed photometric analysis for this system in the $g$ and $i$ bands to derive the galaxy structure and colors. 
Results show that the central object resembles a disk galaxy, surrounded by a ring-like structure 2 times more extended than the central disk. The warped geometry and the presence of bright knots observed along the polar direction, as well as the several debris detected on the NW side with colors comparable to those of the galaxy,  suggest that the polar structure is still forming.  We argue that the wide polar ring/disk is the result of the ongoing disruption of a companion galaxy in the potential of the central object, which is 2-3 times more massive than the accreting galaxy.
 } 
  
   \keywords{Galaxies: photometry; peculiar; structure; formation - Astronomical data bases: Surveys}

\titlerunning{A Forming Wide Polar Ring Galaxy at $z\sim0.05$} 
\authorrunning{Iodice E. et al.}

   \maketitle
%

\section{Introduction}\label{intro}

Polar Ring/Disk Galaxies (PRGs) are multi-spin systems. The polar structure is made by dust and gas that rotates in a perpendicular plane with respect to the stars of the central galaxy \citep[see][as a review]{Iod2014}. A "second event" is invoked in the formation history of PRGs in order to explain the decoupling of the angular momentum. Thus, PRGs are among the best galaxies to study the physics of accretion/interaction mechanisms, the disk formation and the dark halo shape. New deep surveys have shown that multi-spin galaxies  are quite common at high redshift and studying these systems at increasing redshift gives fundamental information on the physical processes at work (i.e. merging and accretion) during the formation of galaxies \citep[see][as a review]{Cons2014}.
In the literature there are few studies on PRGs at high redshifts ($z\ge0.05$), where the dominant component, the central galaxy, is easily visible, while detection of the faint, bluer and dusty polar structure requires deep imaging, with high spatial resolution. 
The most distant kinematically confirmed PRG is at $z\sim0.06$ \citep{Brosch2010}. At higher redshifts ($z\sim 0.06 -1.3$), there are only few PRG candidates to be confirmed \citep{Resh97, Resh07}. The photometric analysis performed on PRGs at $z\ge0.05$ shows that in all of them the polar structure has an almost regular ring-like shape, with some clumps of light due to star forming regions. Up to date there is not a detailed study of a {\it forming} PRG, where the polar structure is still ongoing in an intermediate stage, like the well studied galaxies NGC~3808B and NGC6286 at $z\sim0.02$ \citep{Resh96}. In the new SDSS-based Polar Ring Catalogue (SPRC) compiled by \citet{Moi11}, which reports several new PRG  candidates in a large range of redshift,  only two objects resemble an ongoing interaction to form a PRG at $z\ge0.05$ (SPRC199 and SPRC226), but both of them are neither kinematically confirmed nor studied in detail.
At $z\sim0.02$ a forming polar disk is found in the wall between voids, resulting from the slow accretion of gas \citep{Stan09}: the polar structure is made only by neutral hydrogen (it was detected in the HI emission) and the gas density is still too low to form the stellar counterpart.
In this letter we present the detailed photometric analysis of the background galaxy FCSS J033710.0-354727 at $z\sim0.05$, which is a good candidate to be the first forming wide PRG at this redshift.


\section{Observations and Data Reduction}\label{data}
 
As part of the {\it VST Survey of Elliptical Galaxies in the Southern hemisphere} \citep[VEGAS][see]{Cap2011}, which is a Guaranteed Time Observation survey being performed at the ESO VLT Survey Telescope (VST), we have obtained a mosaic of $1.75 \times 1.59$~degree of the  Deep Field of the Fornax Cluster, around the cD galaxy NGC~1399. Images are collected in the $g$ and $i$ bands, on October 2011. 
VST is a 2.6-m wide field optical survey telescope, located at Cerro Paranal in Chile. 
The VST is currently the largest telescope in the world specially designed for surveying the sky in visible light; it is the ESO work-horse totally dedicated to visible survey programmes. The telescope is a F/5.5 with an alt-azimuth mount, equipped with an active optics system \citep{Schipani2010, Schipani2012}.
VST  is equipped with the wide field camera OmegaCAM, spanning a $1 \times 1$~degree$^2$ field of view, in the optical wavelength range from 0.3 to 1.0 micron \citep{Kui2011}. The mean pixel scale is 0.21~arcsec/pixel.
The region of maximum overlap of all pointings is around the galaxy NGC~1389, where the total integration time is 3 hours in the $g$ band and 1.4 hours in the $i$ band. The average seeing is about 1~arcsec.

The data reduction has been performed with the {\it VST-Tube} imaging pipeline \citep{Grado2012}. From the raw data, it provides fully calibrated images, throughout the following steps: 1)  overscan, bias and flat-field correction; 2) CCD gain equalization and illumination correction; 3) astrometric and photometric calibration, applied before stacking for the final co-added image. For a detailed description of the  data reduction procedure see \citet{Ripepi2014}. 
 In order to perform a deep surface photometry of the extended galaxies in the VEGAS Survey, the VST-Tube pipeline also includes a task to remove  the  background patterns. This will be described  in the forthcoming paper, which is dedicated to the first results of the VEGAS Survey (Capaccioli et al. in preparation). In the present work, since the studied object covers a very small area (its diameter is $D\la 30$~arcsec, see Sec.~\ref{phot}), the background has been estimated locally, in the regions surrounding the galaxy.


\section{The Galaxy FCSS J033710.0-354727: morphology and light distribution}\label{phot}

\begin{figure*}
\centering
\includegraphics[width=17cm]{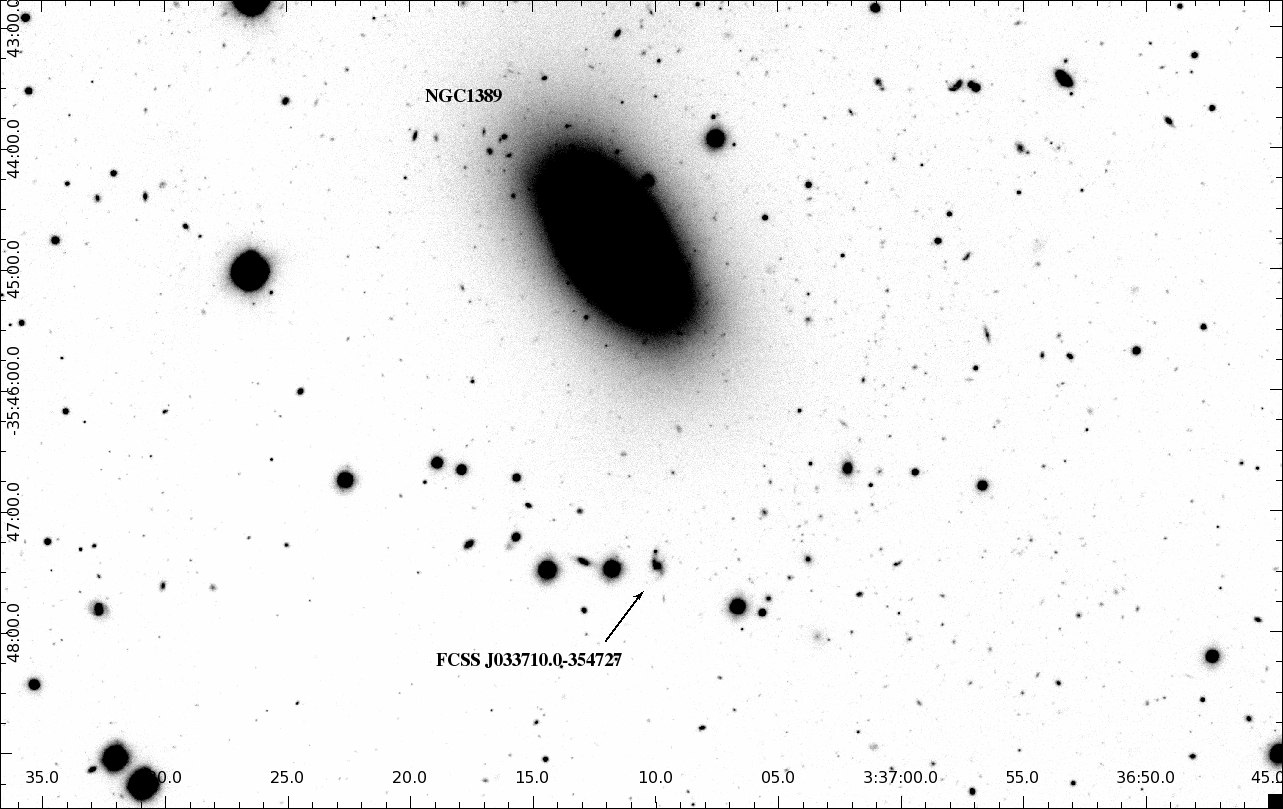} 
\caption{ Field around the bright S0 galaxy NGC~1389, in the $g$-band VST Deep Field of the Fornax cluster. The arrow locates the background galaxy FCSS~J033710.0-354727.} 
\label{field}
\end{figure*}

\begin{figure*}
\centering
\includegraphics[width=17cm]{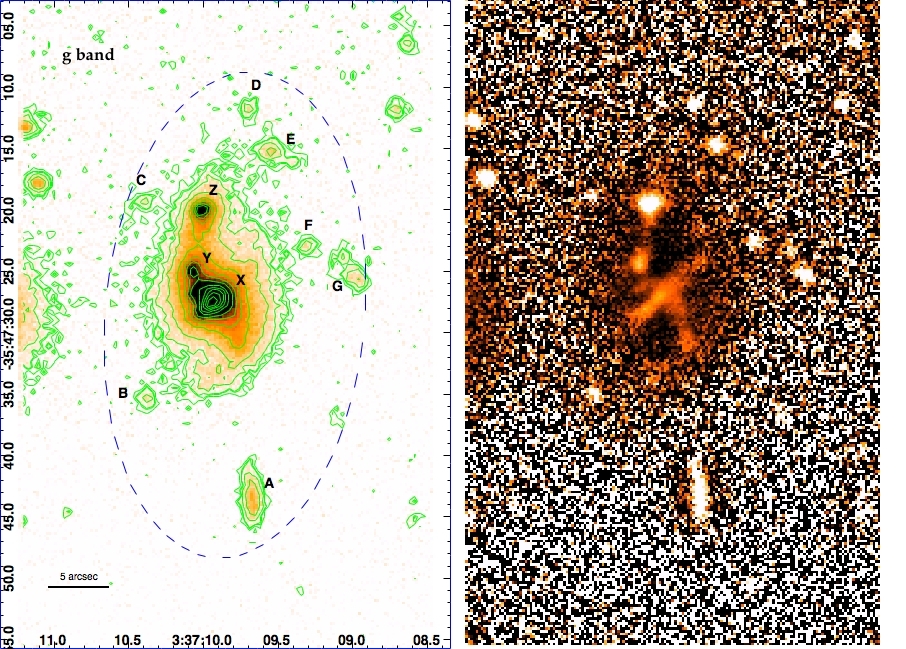} 
\caption{ Left panel - VST image of FCSS~J033710.0-354727 in $g$-band and the isophote contours (green lines). Labels indicate the center of the galaxy, {\bf X}, the two bright knots inside the polar structure,  {\bf Y} and {\bf Z}, and several bright features, from {\bf A} to {\bf G} which are apparently distributed on elliptical orbit (blue dashed line) around the polar direction. Right panel - The {\it high frequency residual images} of FCSS~J033710.0-354727 in the $g$-band (see Sec.~\ref{phot} for details). North is up and East is on the left.} 
\label{PRG}
\end{figure*}

\begin{figure*}
\centering
\includegraphics[width=18cm]{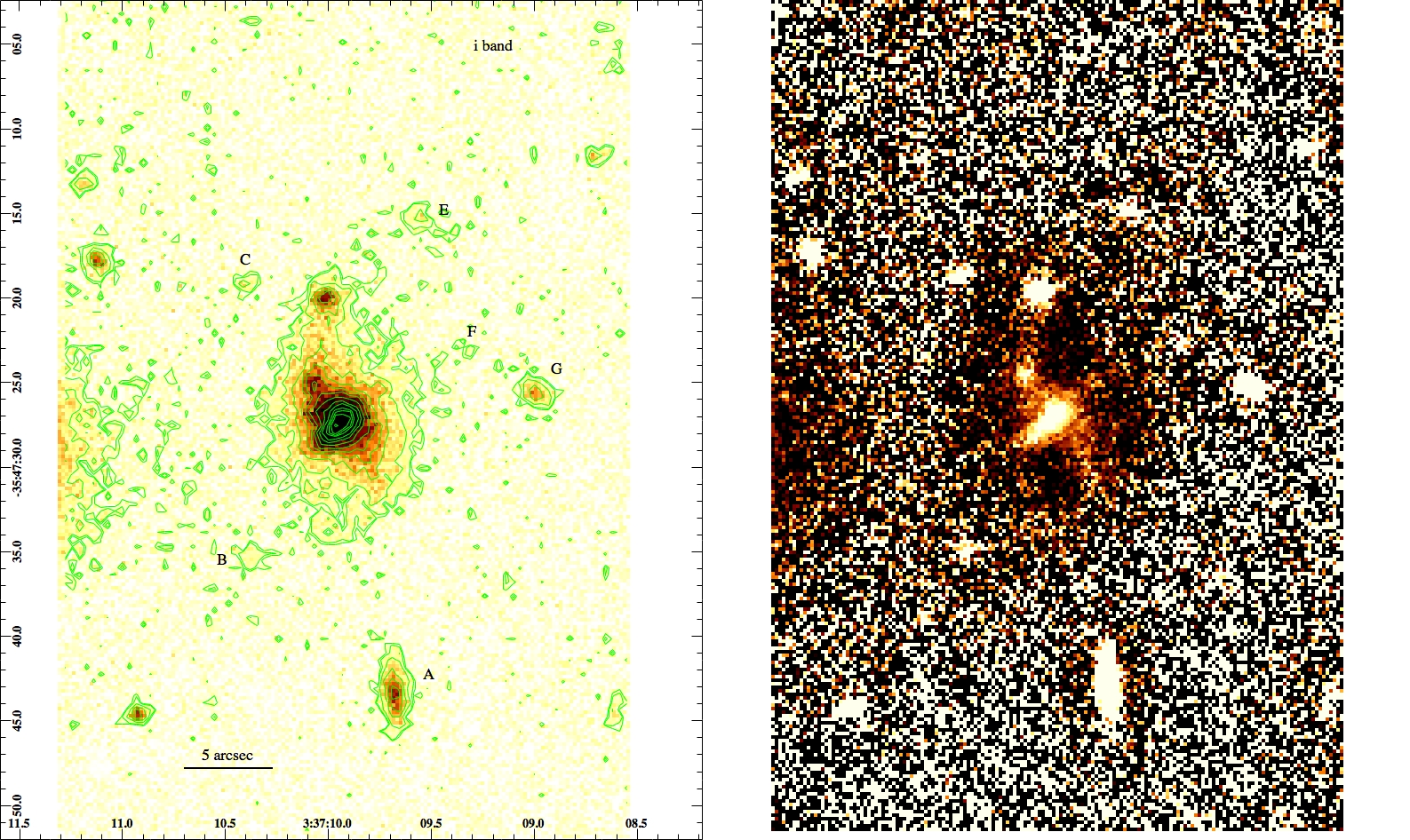} 
\caption{ Left panel - VST image of FCSS~J033710.0-354727 in $i$-band and the isophote contours (green lines). Labels are the same as in Fig.1~\ref{PRG}. Right panel - The {\it high frequency residual images} of FCSS~J033710.0-354727 in the $i$-band. North is up and East is on the left.} 
\label{PRG_i}
\end{figure*}

The background galaxy FCSS~J033710.0-354727 at $z\sim0.051$,  is located South of the bright S0 galaxy NGC~1389 in the Fornax cluster (see Fig.~\ref{field}). It is characterized by a central component, that we named "host galaxy" (HG), surrounded by a warped ring-like structure (see left panel of Fig.~\ref{PRG} and Fig.~\ref{PRG_i}). The main properties of FCSS~J033710.0-354727 are listed in Table\ref{PRG_prop}. 
According to \citet{Brocca97},  inside a distance of about five times its diameter, i.e. $R\le200$~kpc (see Table\ref{PRG_prop}), no companion galaxies at comparable measured redshift  are found around  FCSS~J033710.0-354727, which could be an isolated object.
We have analysed images in the $g$ and $i$ bands to derive the galaxy structure and colors.

\begin{table}
\begin{minipage}[t]{\columnwidth}
\caption{Global properties of FCSS J033710.0-354727.}
\label{PRG_prop}
 \centering
\renewcommand{\footnoterule}{}
\begin{tabular}{lccc}
\hline\hline
Parameter&Value&Ref.\\
\hline
Morphological type&Sc peculiar&NED\footnote{NASA/IPAC Extragalactic Database}\\
R.A. (J2000)           &03h37m09.94s &NED \\
Decl. (J2000)           &-35d47m27.2s&NED \\
Helio. radial velocity & 15447 km/s&NED\\
Redshift         & 0.051 &FCSS\footnote{{\it The Fornax Spectroscopic Survey} \citep{FCSS}}\\
Distance      & 206 Mpc     & \\
scale & 1 kpc/arcsec & \\
Diameters& $3.4\times\ 40$ kpc& this work\\
Magnitudes\footnote{Absolute magnitudes are corrected for both Galactic Extinction and K-correction, while apparent magnitudes take into account only the Galactic Extinction, see text for details.} & & this work\\
 $m_{g}$ & 18.59 mag& \\
$m_{i}$ & 18.03 mag& \\
 $M_{g}$ & -18.00 mag& \\
 $M_{i}$ & -18.48 mag& \\
 & {\it Central galaxy}:&\\ 
 $m_{g}$ & 20.47 mag& \\
 $m_{i}$ & 19.90 mag& \\
 $M_{g}$ &  -16.12 mag& \\
 $M_{i}$ &  -16.61 mag& \\
$g-i$& 0.6 mag & \\
 & {\it Polar structure}:&\\
 $m_{g}$ & 19.09 mag& \\
 $m_{i}$ & 18.53 mag& \\
 $M_{g}$ &  -17.54 mag& \\
 $M_{i}$ & -17.99 mag& \\
$g-i$& $0.56$  mag & \\
\hline
\end{tabular}
\end{minipage}
\end{table}

\begin{figure*}
\centering
\includegraphics[width=17cm]{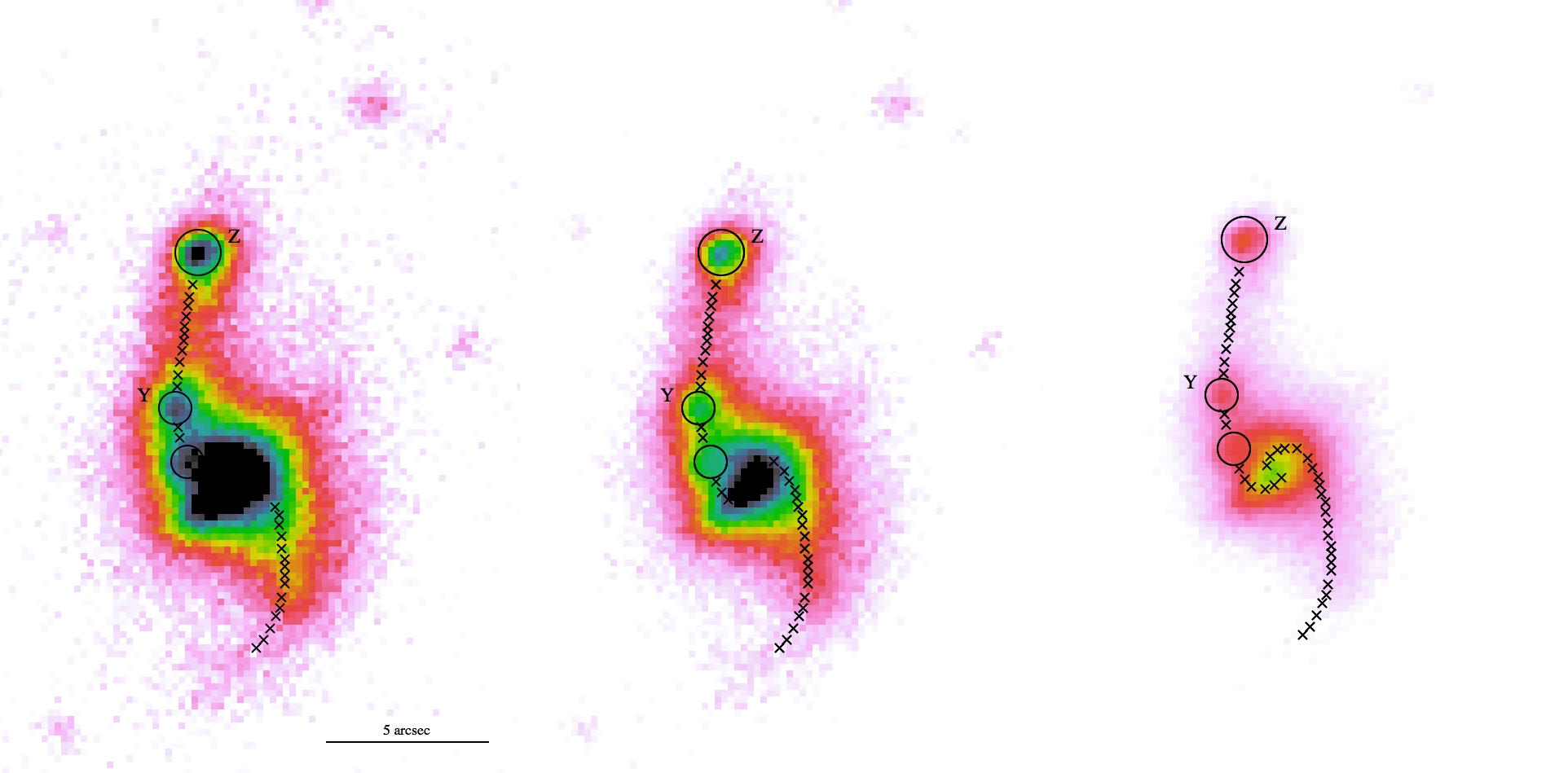} 
\caption{VST images of FCSS~J033710.0-354727 in $g$-band with different contrasts in order 
to emphasize the morphology of the galaxy close to center. The maximum level increases from left to the right. Labels are the same as in Fig.1~\ref{PRG}. The three circles indicates the bright blobs of light along the northern arm of the polar structure. The crosses signature in each panel "follows" the bright path of the polar structure up to the galaxy center (see Sec.~\ref{phot} for details). North is up and East is on the left.} 
\label{PRG_zoom}
\end{figure*}

{\it Morphology -}   The VST image of FCSS~J033710.0-354727 in  $g$ and $i$ bands are shown in the left panels of Fig.~\ref{PRG} and Fig.~\ref{PRG_i}. The center of the galaxy, labelled as {\bf X}, contributes for the most of the light. Towards the North, there are  two other bright knots, labelled as {\bf Y} and {\bf Z}, which are inside the polar structure. At larger distances from the galaxy center, we observe several bright features, which are apparently distributed on elliptical orbit around the polar direction, one of this (labelled as {\bf A}) is likely a disk galaxy. None of the above objects have a redshift measurement. 
We have derived the {\it high frequency residual images} of FCSS~J033710.0-354727 in the $g$ and $i$ bands  (see right panels of Fig.~\ref{PRG} and Fig.~\ref{PRG_i}), as the ratio of the original reduced image with a smoothed\footnote{We used the IRAF task {\small FMEDIAN} to smooth the original reduced image, by adopting a box of $25\times25$~pixels.} one, where each original pixel value is replaced with the median value in a rectangular window.  
The {\it high frequency residual images} shows a disk-like structure along the major axis of the central component.
The polar structure extends up to the galaxy center, crossing it by tracing an "S-shaped" pattern. It is characterized by a very bright peak at the end of the North arm. 
 In Fig.~\ref{PRG_zoom} it is shown the $g$ band image of FCSS~J033710.0-354727 with different contrast:  close to center,  the light of the HG is not symmetric and it is perturbed by the two arms of the polar structure that are approaching the nucleus. In particular, on the NE side there is a third bright knot, which light is smeared in the light coming from the central HG. 
Even if the morphology of FCSS~J033710.0-354727 could be also similar to that of late-type barred galaxy (where the central disk could be the bar with loosely wound and highly-inclined  arms), the qualitative analysis of the structure observed for FCSS~J033710.0-354727 described above suggests to classify this object as PRG. In fact, the two arms of the polar structure cross the center of the galaxy, they do not start from the edge of the central disk/bar.  Moreover, the "accreting loop" observed towards the galaxy center was also observed in other forming polar rings at lower redshifts, such as ESO~474-G26 \citep[see Fig.~14 of][]{Spav12} and VGS31b \citep[see Fig.~4 of][]{Spav13}.

\begin{figure*}
\includegraphics[width=9cm]{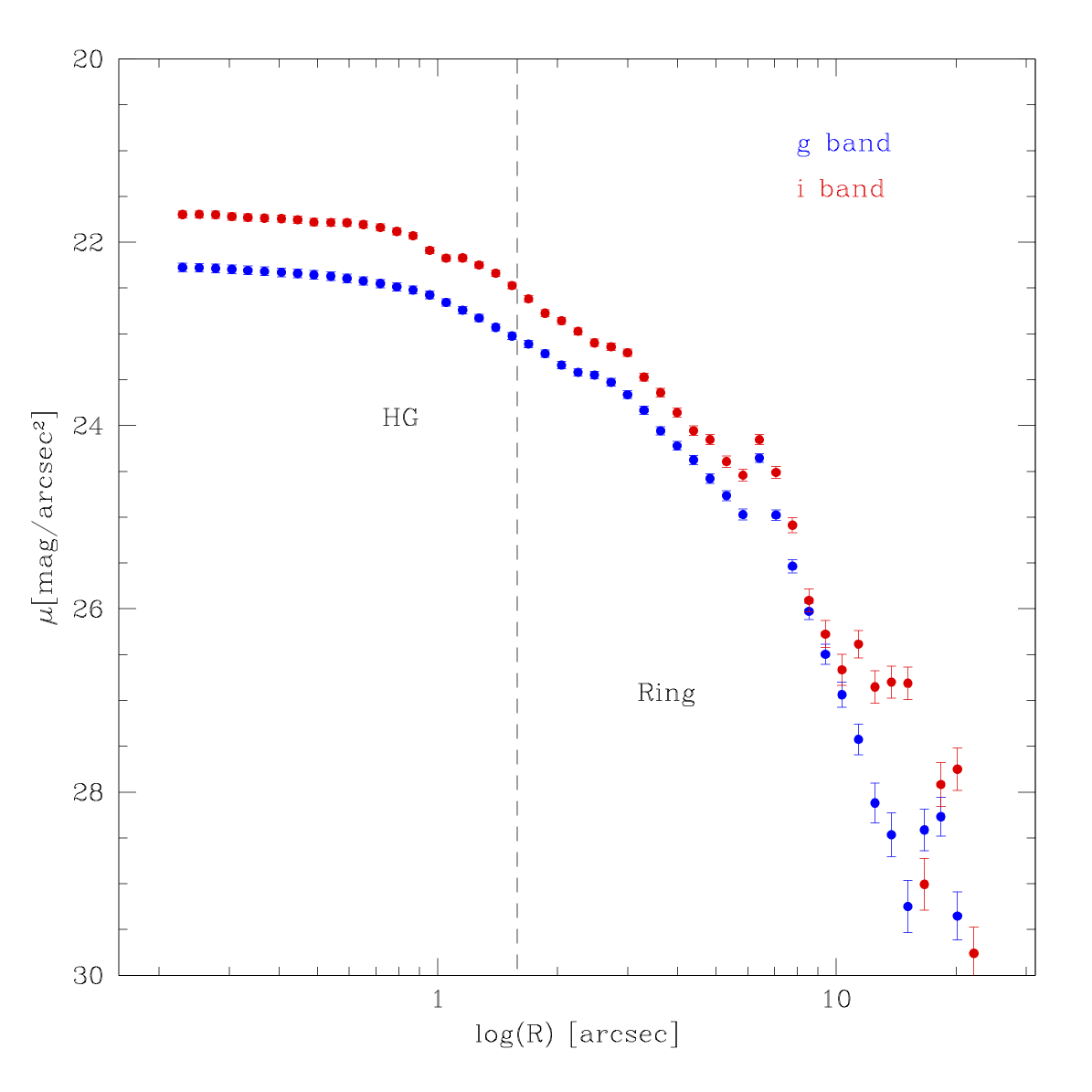} 
\includegraphics[width=9cm]{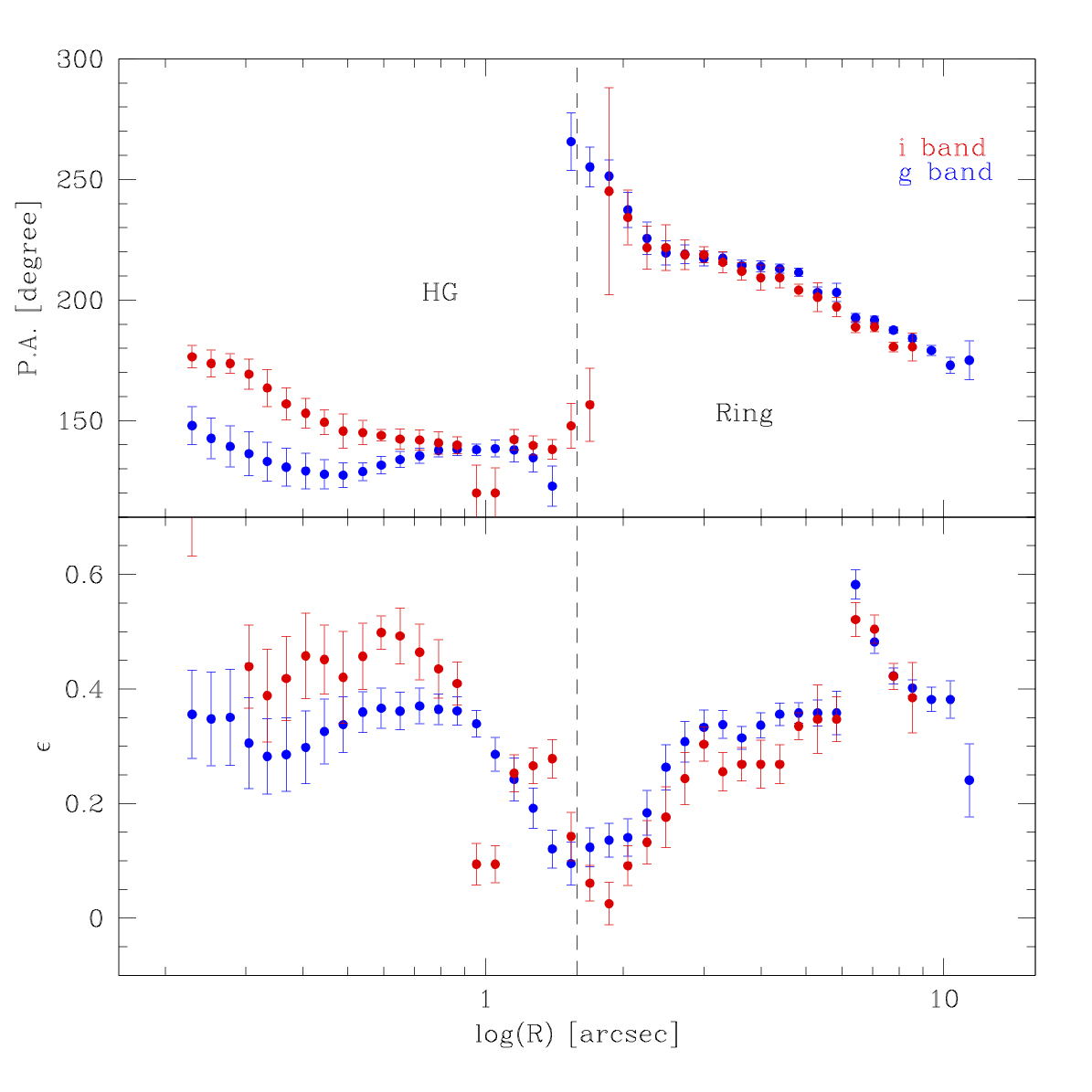} 
\caption{Left panel - Azimuthally averaged surface brightness profiles as function of log(R), derived by the isophote fit. $R$ is the isophote major  axis. Data are for the $g$-band image (blue dots) and $i$-band (red dots). The dashed line delimits the regions where  the main components (HG and polar ring) of the galaxy structure are located. Right panel - Average profiles of P.A. (top panel) and ellipticity (bottom panel) as function of log(R).}
\label{ellipse}
\end{figure*}

{\it Surface photometry -} We used the {\small ELLIPSE} task in IRAF on both $g$ and $i$ band images to perform the isophotal analysis. All the bright features around FCSS~J033710.0-354727, including background objects and bright stars in the field, are masked. The azimuthally averaged surface brightness profiles, the Position Angle (P.A.) and the ellipticity ($\epsilon$) are shown in Fig.~\ref{ellipse}. The limits of the surface photometry presented in this work are  derived as the distance from the center where the galaxy's light blends into the background level, which are found to be 23~arcsec ($\sim$23~kpc) in both $g$ and $i$ band. 
The limiting magnitudes corresponding to the limiting radii given above are $\mu_{g} = 29.3 \pm 0.3$~mag~arcsec$^{-2}$ for the $g$ band, and $\mu_{i} = 27.7 \pm 0.3$~mag~arcsec$^{-2}$ for the $i$ band. 
The error estimates on the magnitudes take the uncertainties on the photometric calibration ($\sim0.02$~mag) and sky subtraction ($\simeq 0.06$~ADU in the $g$ band and $\simeq 0.2$~ADU in the $i$ band) into account.

\begin{figure}
\centering
\includegraphics[width=8cm]{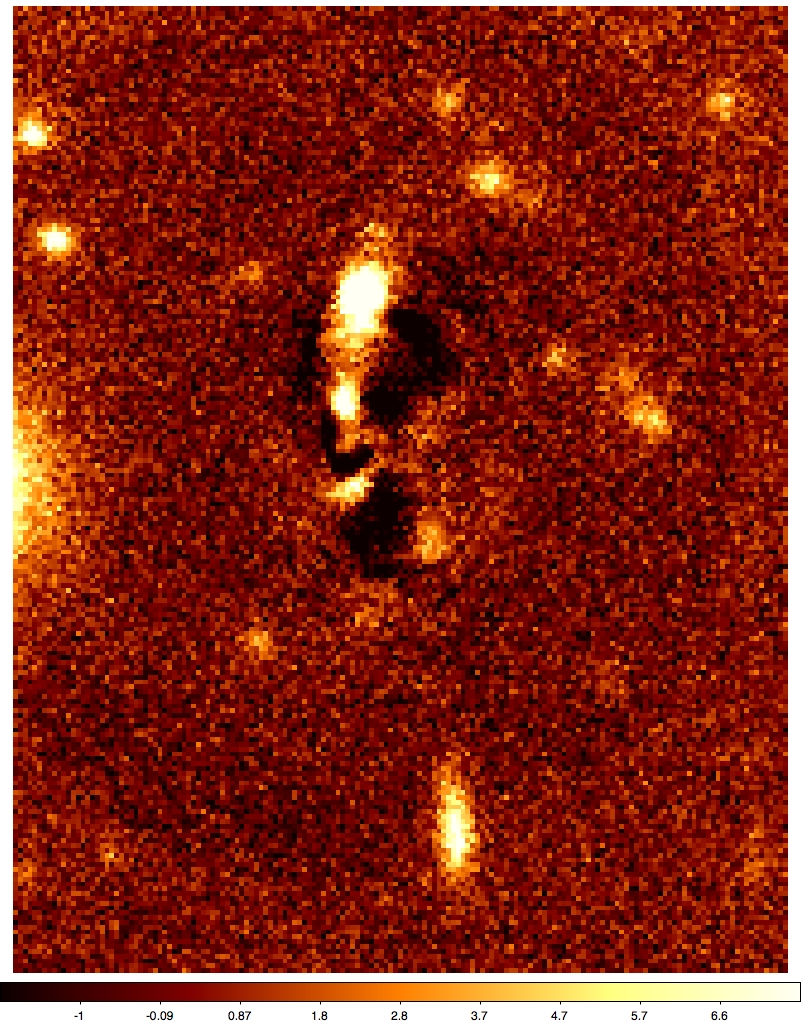} 
\caption{ Residual image obtained by subtracting off the 2D model derived from the fit of isophotes to the original image in the $g$ band. The  image size is $34\times42$~arcsec ($\sim 34\times42$~kpc).}
\label{res}
\end{figure}

The azimuthally averaged surface brightness profiles in the $g$ and $i$ bands (see Fig.~\ref{ellipse}, left panel), are quite smooth on the whole range of radii, except for the two peaks of light at $R\sim 3.5$~arcsec ($\sim3.5$~kpc) and $R\sim 7$~arcsec ($\sim7$~kpc), which corresponds to the bright knots, labeled {\bf Y} and {\bf Z} respectively on Fig.~\ref{PRG} (left panel). 
Both the P.A. and ellipticity profiles (Fig.~\ref{ellipse}), show an abrupt change in the shape at $R\le1.69$~arcsec ($\sim1.7$~kpc): by looking at the isophotes, in the left panel of Fig.~\ref{PRG}, this radius correspond to the "transition" from the HG to polar ring (PR) and set a constraint on the radial extent of the two components, being $R_{HG} \sim 2$~kpc and $R_{PR} \sim 20$~kpc. 
For $R\ge1.69$~arcsec, i.e. along the polar structure major axis, and up to $R \sim 20$~arcsec ($\sim 1.7 - 20$~kpc), a strong twisting is observed, as the P.A. varies of about $70^\circ$. In this range of radii, the flattening increases from  $0.1$ to $0.4$.
 We obtained the 2-dimensional (2D) model  from the fit of the isophotes\footnote{The 2-dimensional model of the fitted isophotes was obtained by using the IRAF task {\small BMODEL}} in the $g$ band  and it has been subtracted off from the original image. The 2D residual is shown in the Fig.~\ref{res}. As already found from the high-frequency residual image (left panels of Fig.~\ref{PRG} and Fig.~\ref{PRG_i}) and from the high-level contrast in the g band (see  Fig.~\ref{PRG_zoom}), along the polar direction an "S-shaped" structure is clearly detectable. It extends from the north to the south crossing the galaxy center,  drawing a spiral loop throughout  the nucleus, and it is characterised by two  bright knots in the northern arm. 

We derived the integrated magnitudes in the $g$ and $i$ bands inside elliptical aperture corresponding to the last fitted isophote for the whole galaxy and for both components (HG and polar structure). Values are corrected or the extinction  within the Milky Way, by using the absorption coefficient  $A_{g}=0.035$ and $A_{i}=0.018$ derived according to   \citet{Schlegel98}. The K-correction  was applied to the absolute magnitudes, where the correction factor in $g$ and $i$ band are $K_g=0.02$ and $K_i=-0.06$ respectively \citep{Chi2010, Chi2012}.

\begin{figure*}
\includegraphics[width=9cm]{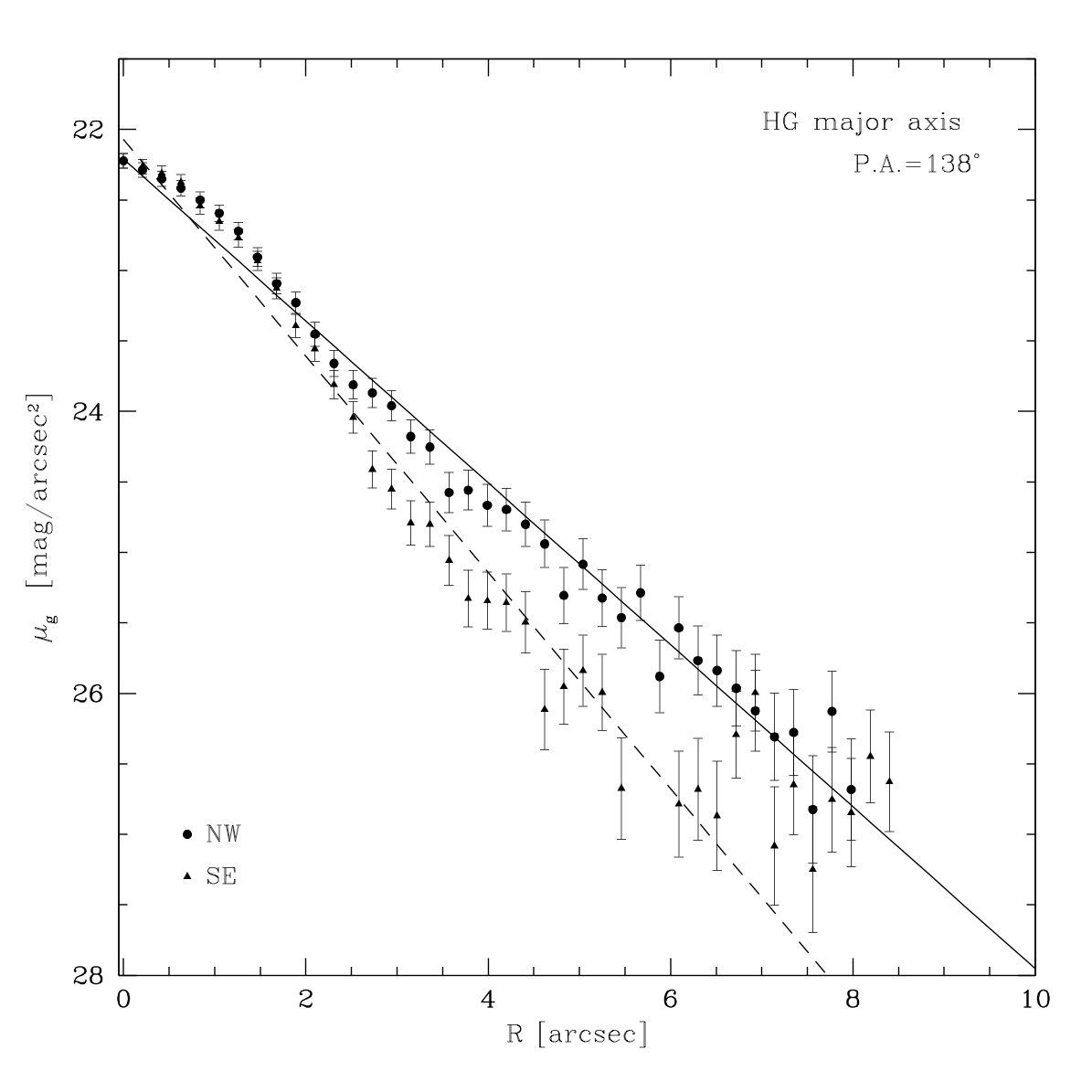} 
\includegraphics[width=9cm]{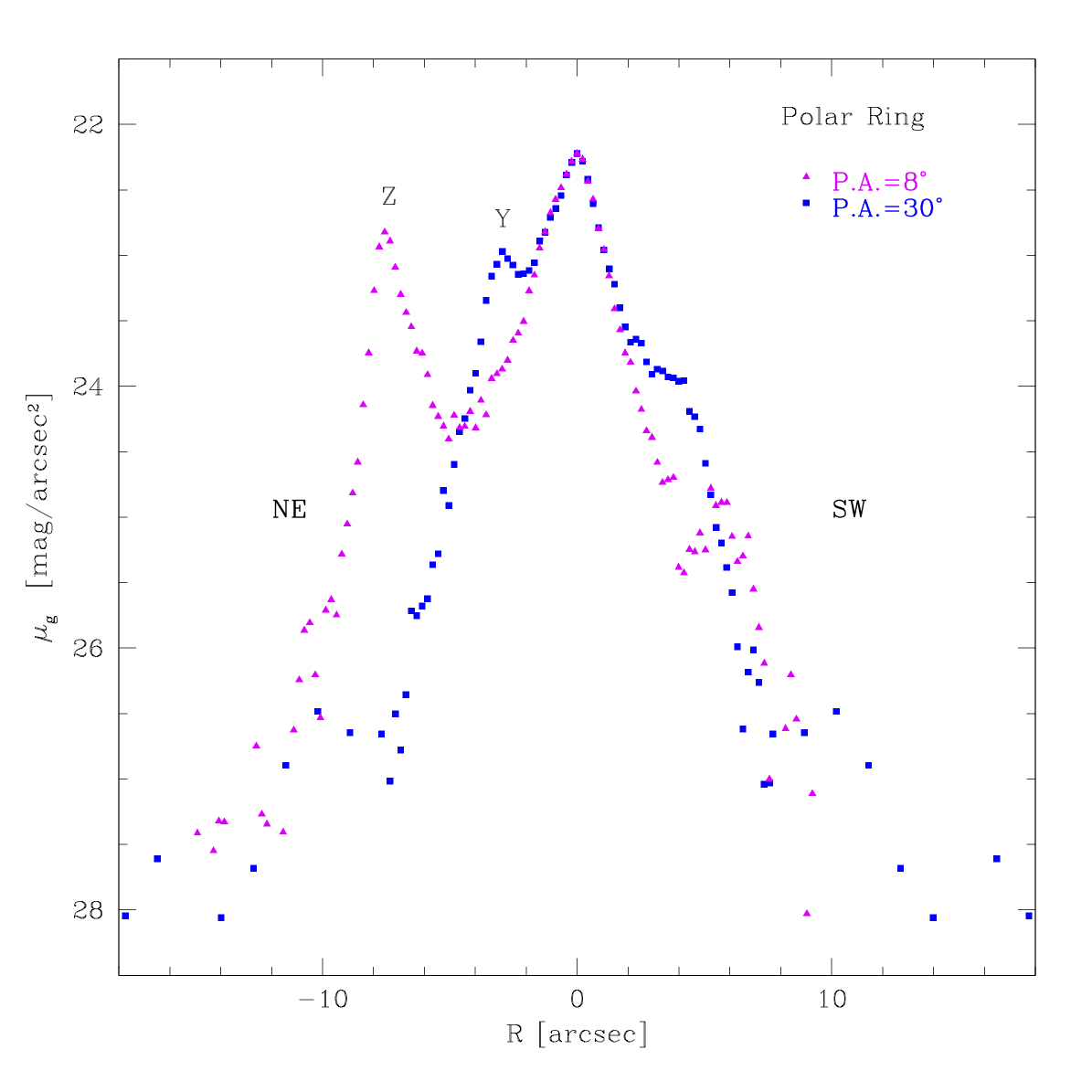} 
\caption{Left panel - Folded light profiles in the $g$ band along the major axis of the central spheroid. The continuos and dashed lines are the results of the exponential fits to the light distribution for NW (circles) and SE (triangles) sides respectively (see Sec.~\ref{phot} for details.)
Right panel -  Light profiles in the $g$ band along the polar structure. 
The  two directions  at $P.A.= 8^\circ$ and $P.A.= 30^\circ$ are chosen to intersect the two bright knots {\bf Y} and {\bf Z}, observed on the north arm, respectively (see also Fig.~\ref{PRG} left panel).}
\label{profili}
\end{figure*}

{\it Light profiles -} We have extracted the light profiles along the the major axis of two main components in FCSS~J033710.0-354727 (see Fig.~\ref{profili}). The P.A.s of these directions are defined from the fit of the isophotes (see right panels of Fig.~\ref{ellipse}).  Due to the "S-shape" of the polar structure, which generates the observed twisting, we adopted  the two directions  at $P.A.= 8^\circ$ and $P.A.= 30^\circ$ that intersect the two bright knots {\bf Y} and {\bf Z}. 

The surface brightness of the central component extends up to about 8~arcsec ($\sim 8$~kpc). It is not symmetric with respect to the center of the galaxy. For $R\ge 2$~arcsec, the NW part of the light profile is brighter than the SE part. We performed a least-square fit of the HG light profiles (NW and SE sides) by using an exponential law given by 
$$\mu(R)= \mu_{0} + 1.086 \times R/r_{h}$$
 where $R$ is the galactocentric distance, $\mu_{0}$ and $r_{h}$ are the central surface brightness and scale length of the disk. The best fit values for the structural parameters found for the NW profile are $\mu^{NW}_{0}=22.21 \pm 0.05$~mag/arcsec$^2$ and $r^{NW}_{h}=1.89 \pm 0.04$~arcsec and for the SE profile are $\mu^{SE}_{0}=22.07 \pm 0.10$~mag/arcsec$^2$ and $r^{SE}_{h}=1.41 \pm 0.05$~arcsec. For $R\le 2$~arcsec there is an excess of light of about 0.2~mag, which is quite symmetric with respect to the center, contrary to what is observed at larger radii. This feature could be related to a small bulge. 

Along the polar structure the surface brightness is twice as extended than that  along the HG, out to $R\sim20$~arcsec ($\sim20$~kpc). The  peak of light corresponding to  {\bf Z} is less than one magnitude fainter than the galaxy center.

\begin{figure}
\centering
\includegraphics[width=8cm]{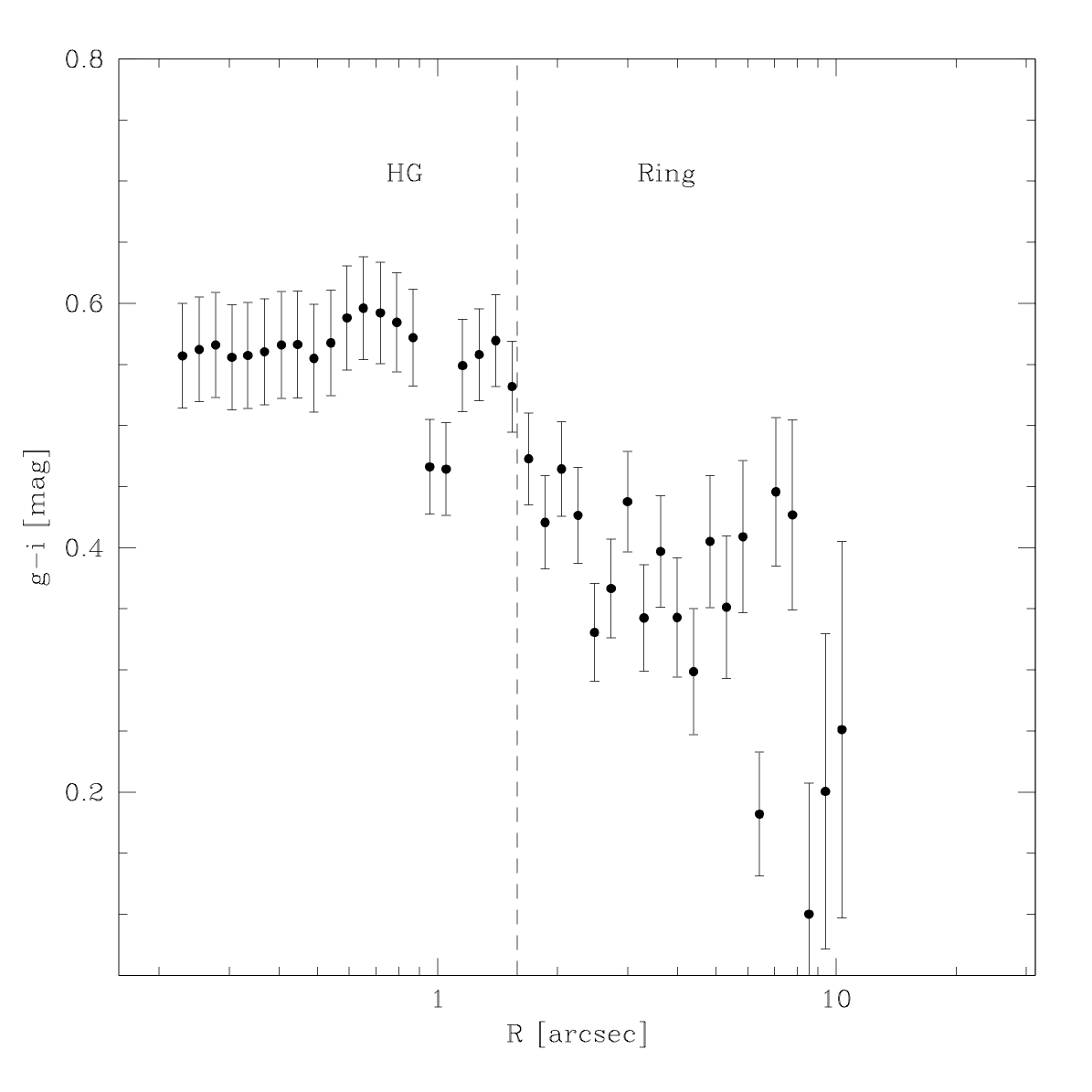} 
\caption{Azimuthally averaged $g-i$ color profile as function of log(R), derived by the isophote fit. $R$ is the isophote major  axis. The dashed line delimits the regions where  the main components (HG and polar ring) of the galaxy structure are located.}
\label{ellipse_col}
\end{figure}

{\it Integrated magnitudes and colors - }  In Fig.~\ref{ellipse_col} it is shown the isophotal $g-i$ color profile. The central spheroid, for $R\le1.6$~arcsec, is redder than the polar structure, having an average color $g-i \sim 0.58$~mag. For $R >1.6$~arcsec, the color profile has a steep gradient towards bluer colors, reaching a value of  $g-i \sim 0.2$~mag at $R\sim 10$~arcsec.

We have derived the integrated magnitudes and $(g-i)$ colors in circular apertures\footnote{The radius of each circular aperture is set to include two times the peak of light.} 
of the center of galaxy, the two bright knots {\bf Y} and {\bf Z}, and of all the bright  objects  around the galaxy (see left panel of Fig.~\ref{PRG}).  Values are listed in Table~\ref{mag}. As already shown from the light profiles, the  bright knot {\bf Z} on the NE side   is only 0.5 magnitude fainter than the galaxy center. 
The center of the galaxy is redder than the two knots on the North side of the polar structure, and it has $(g-i) = 0.55$~mag. 
The closest knot  to the central spheroid, labelled as {\bf Y}, has $(g-i) = 0.35$~mag, while the outer and more luminous knot {\bf Z} of the polar structure has bluer color  $(g-i) = 0.13$~mag.
The integrated colors of the bright  objects observed around FCSS~J033710.0-354727 have a "bimodal" distribution. The galaxy labelled as {\bf A} and the two bright objects {\bf B} and {\bf G} have $(g-i) \ge 1$~mag, while for all the other sources we measured $0.3 \le (g-i) \le 1$~mag. In particular, the objects {\bf D}, {\bf E} and {\bf F}, located on NW region of FCSS J033710.0-354727, have $(g-i)$ colors comparable with the range of colors derived for the polar structure.
From the $g-i$ colors, by using the stellar population synthesis model GISSEL\footnote{\it  Galaxies Isochrone Synthesis Spectral Evolution Library} \citep{Bru03}, we have estimated  the mass-to-light ratio (M/L) for the central galaxy and the polar structure, in order to constrain the total stellar mass for each component. Assuming a simple stellar population, with a solar metallicity, for the central galaxy the models predict an $M/L \sim 0.6$ and for the polar structure  an $M/L \sim 0.3 - 0.6$. From the total magnitudes in the $g$ band (see Table~\ref{mag}), the stellar mass in the HG is $M^{HG} \sim 2 \times\ 10^{8}$~M$_{\odot}$ and in the polar structure is $M^{PR} \sim 1 \times\ 10^{8}$~M$_{\odot}$. These should be considered as lower limits for the total baryonic mass, since we do not have any information about the gas content in this galaxy, which is typically large in PRGs, from  $10^{8}$ to $10^{10}$~M$_{\odot}$, \citep[see][and references therein]{Iod2014}. In particular, since the gas is usually associated to the polar structure, the baryoinc mass of this component in FCSS~J033710.0-354727 could be even higher. Thus, as lower limits for the stellar mass ratio between the central disk and the polar structure we found 3:1 to 2:1.

\begin{table*}
\begin{minipage}[t]{180mm}
\caption{\label{mag}Integrated magnitudes and colors for regions in FCSS J033710.0-354727 in circular apertures.} 
\begin{tabular}{ccccccc}
\hline\hline
Region & $\alpha$ & $\delta$ & R & $m_{g}$ & $m_{i}$  & $g-i$ \\
             &     [sec]   & [arcsec] & [arcsec]  & [mag] & [mag] & [mag]\\
(1)         & (2) & (3) & (4) & (5) & (6) & (7)\\
\hline
\hline
 {\bf X} & 09.93 & 27.15 & 1.05 &   $21.14\pm 0.02$ &  $20.61\pm 0.02$ & $0.53\pm0.04$\\
 {\bf Y} & 10.07 & 24.71 & 1.05 &  $21.98\pm0.03$ & $21.63\pm0.03$ & $0.35\pm0.06$\\
 {\bf Z} & 09.94 & 20.17 & 1.47 &  $21.64\pm0.02$ & $21.53\pm0.03$ & $0.11\pm0.05$\\
 {\bf A} & 09.67 & 43.39 & 2.31 &  $22.57\pm0.04$ & $21.30\pm0.02$ & $1.27\pm0.05$\\
 {\bf B} & 10.38 & 35.53 & 1.89 &  $23.68\pm0.07$ & $22.65\pm0.06$ & $1.03\pm0.13$\\
 {\bf C} & 10.41 & 19.29 & 1.47 &  $23.95\pm0.08$ & $23.01\pm0.07$ & $0.94\pm0.15$\\
 {\bf D} & 09.69 & 11.61 & 1.89 &  $23.75\pm0.08$ & $23.49\pm0.12$ & $0.26\pm0.2$\\
 {\bf E} & 09.55 & 15.28 & 1.89 &  $23.16\pm0.06$ & $22.48\pm0.06$ & $0.68\pm0.12$\\
 {\bf F} & 09.30 & 23.13 & 1.47 &  $24.00\pm0.08$ & $23.41\pm0.10$ & $0.60\pm0.2$\\
 {\bf G} & 09.00 & 24.79 & 2.31 &  $23.00\pm0.05$ & $21.93\pm0.04$ & $1.07\pm0.09$\\
\hline
\end{tabular}
\end{minipage}
\smallskip

{\em Col.~1}: Region of FCSS J033710.0-354727 labelled in Fig.~\ref{PRG}.  {\em Col.~2} and {\em Col.~3}: Celestial coordinates  of the the center of each circular aperture, see left panel of Fig.~\ref{PRG} for reference. {\em Col.~4} Radius  of the circular aperture in arcsec. {\em Col.~5} and {\em Col.~6}: Integrated magnitudes in the $g$ and $i$ bands corrected for the Galactic Extinction. {\em Col.~7} Integrated $g-i$ color.
\end{table*}


\section{Results: the Ongoing Formation of a Wide Polar Ring/Disk}

 The deep exposures in the $g$ and $i$ bands, combined with the high angular resolution of the OmegaCAM at VST, allow us to carry out the first detailed photometric analysis of the background galaxy FCSS~J033710.0-354727 at $z\sim0.05$, in the field of the Fornax cluster. 

 Main results obtained in the present work (see Sec.\ref{phot}) show that:

\begin{itemize}
\item the system is  characterized by a central component, surrounded by a warped ring-like structure (see Fig.~\ref{PRG}  and Fig.~\ref{PRG_zoom}.);
\item the central component is a disk with an exponential surface brightness profile;
\item the polar structure is 2 times more extended than the central disk. It  crosses the galaxy center, along the North-South direction,  drawing a spiral loop throughout  the nucleus.  It is characterised by two bright knots, one of them ({\bf Z}) having an almost comparable luminosity  to that of the galaxy center;
\item the central galaxy  is redder ($g-i \sim 0.55$~mag) than the polar structure ($g-i \sim 0.13 - 0.4$~mag); 
\item the integrated colors of  the bright objects detected around FCSS~J033710.0-354727, have a "bimodal" distribution. Only those located on the NW region have $(g-i)$ colors comparable with those derived for the polar structure (see Table~\ref{mag}), thus they can be considered as features "related" to the galaxy.
\end{itemize}

The new observations and analysis are in favor of the classification of this galaxy as PRG. In particular, the whole morphology and the large polar extension suggests that this system can be considered as good candidate for a wide polar ring/disk galaxy, like NGC~4650A \citep{Iod02}. In FCSS~J033710.0-354727, the warped geometry of the polar structure and the presence of bright knots along its light distribution, as well as the several debris observed on the NW side with similar colors, suggest that the polar structure is still forming. 
Given the high luminosity, comparable with that of the galaxy center, the two knots {\bf Y} and {\bf Z} (see Fig.~\ref{PRG})  could be the remnant of a companion galaxy that is disrupting in the potential of the central disk, which is 2-3 times more massive than the accreting object. This mechanism, i.e. the tidal accretion of material from outside, is one of the possible formation scenario proposed for PRGs \citep[see][as review]{Combes2014}. In this framework, taking into account the small stellar mass ratio between the central disk and the polar structure (3:1 - 2:1), and the high inclination of the accreting material, numerical simulations \citep{Bou03} are able to form a massive and extended polar ring/disk as observed in FCSS~J033710.0-354727. By comparing the snapshots of the simulation with the observed morphology for FCSS~J033710.0-354727, we suggest that this kind of gravitational interaction is still in act, thus  we are looking at the intermediate stage of  the PRG formation, at the epoch of about 2 Gyr.

Alternatively, FCSS~J033710.0-354727 might also be a post-merger system, where two big galaxies have already coalesced into the single system, with two tidal tails around it and an accumulation of mass at the tip of the northern tail. Something like NGC~7252  \citep{Hibb95}. 
One  observational fact in favour of the tidal accretion process, instead of the major merger, is the large mass of the the bright knot {\bf Z}, not typically found in tidal tail, and the "S-shape" feature that connects the outer parts of polar structure to the galaxy center (see right panel of Fig.~\ref{PRG}), which is more similar to an  accreting loop rather than a tidal tail remnants.
Kinematic measurements are needed, not only to confirm  FCSS~J033710.0-354727 as PRG, but also to discriminate between the two formation mechanisms. In particular, if this galaxy is the remnant of a major merger between two disk galaxies we expect to found the central object dominated by random motions \citep{Bou03}.

In conclusion, FCSS~J033710.0-354727 is  a peculiar system at $z\sim0.05$ resulting from the interaction of two massive galaxies, which could form a wide polar ring/disk galaxy. 
As discussed in Sec.~\ref{intro}, the few PRGs studied at $z \ge 0.05$ show a well-formed polar structure,  without any tail or ripple that could suggest an ongoing interaction event.
The main result of this work is that, up to date, FCSS~J033710.0-354727 is the most distant  {\it PRGs in formation} for which a detailed photometric analysis has been performed: the intermediate stage "captured" for this object allows us to derive the important constraint on the mass ratio of the two interacting galaxies, which is hard to estimate, or quite uncertain, in a final remnant galaxy.


\begin{acknowledgements}
We are very grateful to the anonymous referee  for his/her comments and
suggestions which helped us to improve and clarify our work. 
This work  was supported by the PRIN-INAF "Galaxy Evolution with the VLT Survey Telescope (VST)" (PI A. Grado). M. Cantiello acknowledges support from PO FSE Abruzzo 2007-2013 (PO 2012/2013).
\end{acknowledgements}

\bibliographystyle{aa}
\bibliography{bibliografia.bib}

\begin{thebibliography}{26}
\expandafter\ifx\csname natexlab\endcsname\relax\def\natexlab#1{#1}\fi

\bibitem[{{Bournaud} \& {Combes}(2003)}]{Bou03}
{Bournaud}, F. \& {Combes}, F. 2003, A\&A, 401, 817

\bibitem[{{Brocca} {et~al.}(1997){Brocca}, {Bettoni}, \& {Galletta}}]{Brocca97}
{Brocca}, C., {Bettoni}, D., \& {Galletta}, G. 1997, A\&A, 326, 907

\bibitem[{{Brosch} {et~al.}(2010){Brosch}, {Kniazev}, {Moiseev}, \&
  {Pustilnik}}]{Brosch2010}
{Brosch}, N., {Kniazev}, A.~Y., {Moiseev}, A., \& {Pustilnik}, S.~A. 2010,
  \mnras, 401, 2067

\bibitem[{{Bruzual} \& {Charlot}(2003)}]{Bru03}
{Bruzual}, G. \& {Charlot}, S. 2003, MNRAS, 344, 1000

\bibitem[{{Capaccioli} \& {Schipani}(2011)}]{Cap2011}
{Capaccioli}, M. \& {Schipani}, P. 2011, The Messenger, 146, 2

\bibitem[{{Chilingarian} {et~al.}(2010){Chilingarian}, {Melchior}, \&
  {Zolotukhin}}]{Chi2010}
{Chilingarian}, I.~V., {Melchior}, A.-L., \& {Zolotukhin}, I.~Y. 2010, \mnras,
  405, 1409

\bibitem[{{Chilingarian} \& {Zolotukhin}(2012)}]{Chi2012}
{Chilingarian}, I.~V. \& {Zolotukhin}, I.~Y. 2012, \mnras, 419, 1727

\bibitem[{{Combes}(2014)}]{Combes2014}
{Combes}, F. 2014, in Astronomical Society of the Pacific Conference Series,
  Vol. 486, Multi-Spin Galaxies, ASP Conference Series, ed. E.~{Iodice} \&
  E.~M. {Corsini}, 207

\bibitem[{{Conselice}(2014)}]{Cons2014}
{Conselice}, C.~J. 2014, in Astronomical Society of the Pacific Conference
  Series, Vol. 486, Multi-Spin Galaxies, ASP Conference Series, ed. E.~{Iodice}
  \& E.~M. {Corsini}, 85

\bibitem[{{Drinkwater} {et~al.}(2000){Drinkwater}, {Phillipps}, {Jones},
  {Gregg}, {Deady}, {Davies}, {Parker}, {Sadler}, \& {Smith}}]{FCSS}
{Drinkwater}, M.~J., {Phillipps}, S., {Jones}, J.~B., {et~al.} 2000, \aap, 355,
  900

\bibitem[{{Grado} {et~al.}(2012){Grado}, {Capaccioli}, {Limatola}, \&
  {Getman}}]{Grado2012}
{Grado}, A., {Capaccioli}, M., {Limatola}, L., \& {Getman}, F. 2012, Memorie
  della Societa Astronomica Italiana Supplementi, 19, 362

\bibitem[{{Hibbard} \& {Mihos}(1995)}]{Hibb95}
{Hibbard}, J.~E. \& {Mihos}, J.~C. 1995, \aj, 110, 140

\bibitem[{{Iodice}(2014)}]{Iod2014}
{Iodice}, E. 2014, in Astronomical Society of the Pacific Conference Series,
  Vol. 486, Astronomical Society of the Pacific Conference Series, ed.
  E.~{Iodice} \& E.~M. {Corsini}, 39

\bibitem[{{Iodice} {et~al.}(2002){Iodice}, {Arnaboldi}, {De Lucia},
  {Gallagher}, {Sparke}, \& {Freeman}}]{Iod02}
{Iodice}, E., {Arnaboldi}, M., {De Lucia}, G., {et~al.} 2002, AJ, 123, 195

\bibitem[{{Kuijken}(2011)}]{Kui2011}
{Kuijken}, K. 2011, The Messenger, 146, 8

\bibitem[{{Moiseev} {et~al.}(2011){Moiseev}, {Smirnova}, {Smirnova}, \&
  {Reshetnikov}}]{Moi11}
{Moiseev}, A.~V., {Smirnova}, K.~I., {Smirnova}, A.~A., \& {Reshetnikov}, V.~P.
  2011, MNRAS, 418, 244

\bibitem[{{Reshetnikov}(1997)}]{Resh97}
{Reshetnikov}, V.~P. 1997, A\&A, 321, 749

\bibitem[{{Reshetnikov} \& {Dettmar}(2007)}]{Resh07}
{Reshetnikov}, V.~P. \& {Dettmar}, R.-J. 2007, Astronomy Letters, 33, 222

\bibitem[{{Reshetnikov} {et~al.}(1996){Reshetnikov}, {Hagen-Thorn}, \&
  {Yakovleva}}]{Resh96}
{Reshetnikov}, V.~P., {Hagen-Thorn}, V.~A., \& {Yakovleva}, V.~A. 1996, \aap,
  314, 729

\bibitem[{{Ripepi} {et~al.}(2014){Ripepi}, {Cignoni}, {Tosi}, {Marconi},
  {Musella}, {Grado}, {Limatola}, {Clementini}, {Brocato}, {Cantiello},
  {Capaccioli}, {Cappellaro}, {Cioni}, {Cusano}, {Dall'Ora}, {Gallagher},
  {Grebel}, {Nota}, {Palla}, {Romano}, {Raimondo}, {Sabbi}, {Getman},
  {Napolitano}, {Schipani}, \& {Zaggia}}]{Ripepi2014}
{Ripepi}, V., {Cignoni}, M., {Tosi}, M., {et~al.} 2014, \mnras, 442, 1897

\bibitem[{{Schipani} {et~al.}(2010){Schipani}, {D'Orsi}, {Ferragina}, {Fierro},
  {Marty}, {Molfese}, \& {Perrotta}}]{Schipani2010}
{Schipani}, P., {D'Orsi}, S., {Ferragina}, L., {et~al.} 2010, \ao, 49, 1234

\bibitem[{{Schipani} {et~al.}(2012){Schipani}, {Noethe}, {Arcidiacono},
  {Argomedo}, {Dall'Ora}, {D'Orsi}, {Farinato}, {Magrin}, {Marty}, {Ragazzoni},
  \& {Umbriaco}}]{Schipani2012}
{Schipani}, P., {Noethe}, L., {Arcidiacono}, C., {et~al.} 2012, Journal of the
  Optical Society of America A, 29, 1359

\bibitem[{{Schlegel} {et~al.}(1998){Schlegel}, {Finkbeiner}, \&
  {Davis}}]{Schlegel98}
{Schlegel}, D.~J., {Finkbeiner}, D.~P., \& {Davis}, M. 1998, ApJ, 500, 525

\bibitem[{{Spavone} \& {Iodice}(2013)}]{Spav13}
{Spavone}, M. \& {Iodice}, E. 2013, mnras, 434, 3310

\bibitem[{{Spavone} {et~al.}(2012){Spavone}, {Iodice}, {Bettoni}, {Galletta},
  {Mazzei}, \& {Reshetnikov}}]{Spav12}
{Spavone}, M., {Iodice}, E., {Bettoni}, D., {et~al.} 2012, mnras, 426, 2003

\bibitem[{{Stanonik} {et~al.}(2009){Stanonik}, {Platen}, {Arag{\'o}n-Calvo},
  {van Gorkom}, {van de Weygaert}, {van der Hulst}, \& {Peebles}}]{Stan09}
{Stanonik}, K., {Platen}, E., {Arag{\'o}n-Calvo}, M.~A., {et~al.} 2009, \apjl,
  696, L6

\end{thebibliography}



\end{document}